\documentclass[final, twocolumn]{elsarticle}

\usepackage{amssymb}

\newtheorem{theorem}{\bf Theorem}

\newtheorem{condition}{\bf Condition}

\journal{Ad Hoc Networks}

\begin{document}

\emergencystretch 3em
\raggedbottom

\begin{frontmatter}



\title{Distributed Multiple Access with A General Link Layer Channel}


\author{
  {Yanru Tang, Faeze Heydaryan, and Jie Luo}
}

\address{
{Electrical \& Computer Engineering Department} \\
  {Colorado State University, Fort Collins, CO 80523} \\
  {Email: \{yrtang, faezeh66, rockey\}@colostate.edu} }

\begin{abstract}
This paper investigates the problem of distributed medium access control in a time slotted wireless multiple access network with an unknown finite number of homogeneous users. Assume that each user has a single transmission option. In each time slot, a user chooses either to idle or to transmit a packet. Under a general channel model, a distributed medium access control framework is proposed to adapt transmission probabilities of all users to a value that maximizes an arbitrarily chosen symmetric network utility. Probability target of each user in the proposed algorithm is calculated based upon a channel contention measure, which is defined as the success probability of a virtual packet. It is shown that the proposed algorithm falls into the classical stochastic approximation framework with guaranteed convergence when the contention measure can be directly obtained from the receiver. On the other hand, when the contention measure is not directly available, computer simulations show that a revised medium access control algorithm can still help the system to converge to the same designed equilibrium.
\end{abstract}


\end{frontmatter}


\section{Introduction}
In distributed communication when users access the channel opportunistically, packet collisions are often unavoidable. When communication optimization cannot be achieved fully at the physical layer, data link layer must get involved in communication adaptation. To support efficient data link layer adaptation, great efforts have been made to investigate the problem of distributed medium access control (MAC). Distributed MAC protocols can generally be categorized into non-adaptive ALOHA protocols \cite{ref Abramson70}, splitting algorithms \cite{ref Bertsekas92}\cite{ref Qin04} and back-off approaches \cite{ref Hajek85}\cite{ref Bianchi00}. ALOHA protocols are often used to investigate fundamental limits, such as the throughput and the stability regions, of a network \cite{ref Rao88}\cite{ref Luo06}. In splitting algorithms such as the FCFS algorithm \cite{ref Bertsekas92}, users maintain a common virtual interval of their random identity values. Users partition the interval based upon a sequence of channel feedback messages and determine their transmission schedules accordingly. While splitting algorithms can often achieve a relatively high system throughput, their function depends on the assumptions of instant availability of noiseless channel feedback and correct reception of feedback sequences. Unfortunately, both of these conditions can be violated in a wireless environment. Theoretical analysis of a splitting algorithm can be extremely challenging, especially when wireless-related factors such as channel noise, feedback error, and transmission delay are taken into account. Back-off algorithms \cite{ref Hajek82}\cite{ref Hajek85}\cite{ref Bianchi00}, on the other hand, have proven to enjoy more trackable analysis. In back-off algorithms such as the 802.11 DCF protocol, according to packet availability, each user transmits its packets with an associated probability parameter. A user should decrease its transmission probability in response to a packet collision (or transmission failure) event, and increase its transmission probability in response to a transmission success event. Distributed probability adaptation in a back-off algorithm often falls into the stochastic approximation framework \cite{ref Hajek82}\cite{ref Hajek85}, with rigorously developed mathematical and statistical tools available for its convergence and performance analysis. It is well known that convergence proof of these algorithms often hold in the existence of measurement noise and feedback delay \cite{ref Kushner97}. Practical back-off algorithms can also be analyzed using Markov models to characterize the impact of discrete probability updates \cite{ref Bianchi00}.

In \cite{ref Hajek85}, a stochastic approximation model was proposed for distributed networking over a collision channel with an unknown finite number of users, each having a saturated message queue. By setting the transmission probability target of each user as a function of the channel idling probability, it was shown that the system can be designed to converge to a unique stable equilibrium. In the case of throughput maximization with homogeneous users, it was proposed that idling probability of the channel should be controlled toward the asymptotically optimal value of $1/e$. This is similar to the proposal of controlling the total traffic level toward $1$, as discussed in \cite{ref Hajek82} using a stochastic approximation framework for a system with an infinite number of users. Most of the existing analysis of the splitting and the back-off algorithms either assume a throughput optimization objective and/or a simple collision channel model. While significant research efforts have been made to revise collision resolution algorithms to incorporate wireless-related physical layer properties, such as capture effect \cite{ref Lau92} and multi-packet reception \cite{ref Ghez88}, not much progress has been reported since the 1980s on integrating these extensions with the insightful stochastic approximation-based frameworks, such as those introduced in \cite{ref Hajek82}\cite{ref Hajek85}.

Recently, an extended channel coding theory was developed in \cite{ref Luo12}\cite{ref Wang12}\cite{ref Luo15} for physical layer distributed communication that features opportunistic channel access and packet collision. The new coding theory enabled the derivation of fundamental limits of distributed communication systems. It also supported the derivation of a link layer channel model based upon the physical layer channel and coding parameters of the data packets. This motivated the investigation on the impact of a general link layer channel model to the design and optimization of collision resolution algorithms.

In this paper, we consider the problem of distributed utility optimization in a wireless multiple access network with an unknown finite number of homogeneous users. Assume that each user is backlogged with a saturated message queue. The link layer multiple access channel is generally modeled by two sets of channel parameters, detailed in the paper. Given the channel model, we propose a distributed MAC framework for each user to adapt its transmission probability according to a channel contention measure defined as the success probability of a virtual packet. We show that the proposed MAC algorithm falls into the classical stochastic approximation framework with guaranteed convergence, if the underlying ordinary differential equation (ODE) has a unique equilibrium and two key monotonicity conditions are satisfied. Without knowing the number of users, we show that one can develop the MAC algorithm to satisfy the required conditions and to place the unique equilibrium at a point that is not far from optimal with respect to a chosen utility. Our work extends the basic framework of \cite{ref Hajek85} from a simple collision channel model to a general link-layer channel. Such extension is enabled by the following key ideas. First, as opposed to measuring contention level of the channel using a locally observable variable such as the channel idling probability \cite{ref Hajek85}, we measure channel contention level using the success probability of a carefully designed virtual packet. Coding parameters of the virtual packet affect the optimality of the MAC algorithm through a set of channel parameters that may need to be derived using the distributed channel coding theorems presented in \cite{ref Luo12}\cite{ref Wang12}\cite{ref Luo15}. Second, with the help of the channel contention measure and two key monotonicity properties, we show that each user can first estimate the unknown number of users, and then set its transmission probability target as a function of the estimated number of users. Compared with the approach of maintaining channel contention at a more or less fixed level for all values of number of users, as suggested in \cite{ref Hajek85}, the MAC algorithm to be proposed can help the system to achieve a performance closer to optimal especially when the number of users is not large in value.

The rest of the paper is organized as follows. In Section \ref{SectionII}, we present a stochastic approximation framework for a class of distributed MAC algorithms. While the framework and its convergence results are quite standard in the stochastic approximation literature, they characterize the key conditions for guaranteed convergence to a unique system equilibrium. Within the framework, the research objectives become to design the system to satisfy the convergence conditions, and to place the unique equilibrium at a point that maximizes a chosen network utility. We propose to measure contention level of the channel using the success probability of a carefully designed virtual packet, and require that users should derive a common transmission probability target as a function of the common channel contention measure. Such an approach guarantees that transmission probabilities of all users at any system equilibrium must be identical. In Section \ref{SectionIII}, under the assumption that channel contention measure can be directly fed back by the receiver to the transmitters, we propose a distributed MAC algorithm to adapt the transmission probabilities of the users to lead the actual channel contention level toward a predetermined theoretical value. Convergence of the proposed MAC algorithm is proven with the help of two key monotonicity properties. In Section \ref{SectionIV}, we consider the more practical case when each user only knows the conditional success probability of its own packets. A two step approach is proposed for each user to interpret the channel contention measure and to adapt its transmission probability accordingly. Simulation results are provided in Section \ref{SectionV} to demonstrate both the optimality and the convergence properties of the proposed MAC algorithms under various system settings.

To help reading the technical contents of the paper, we summarize the definitions of a list of key variables below.

\begin{enumerate}
\item[] \textbf{Definitions of Key Variables}
\item[] $\{C_{rj}\}:$ real channel parameter set. $C_{rj}$ is the conditional success probability of a real packet should it be transmitted in parallel with $j$ other real packets.
\item[] $\{C_{vj}\}:$ virtual channel parameter set. $C_{vj}$ is the success probability of the virtual packet should it be transmitted in parallel with $j$ real packets.
\item[] $J_{\epsilon_v}:$ $\mathop{\arg\min}_j C_{vj}>C_{v(j+1)}+\epsilon_v$.\item[] $K:$ actual number of users.
\item[] $\hat{K}$: estimated number of users.
\item[] $\mbox{\boldmath $p$}:$ transmission probability vector of all users.
\item[] $\tilde{\mbox{\boldmath $p$}}:$ transmission probability vector target computed using noisy measurements.
\item[] $\hat{\mbox{\boldmath $p$}}:$ theoretical transmission probability vector target computed using noiseless measurements.
\item[] $\mbox{\boldmath $p$}^*:$ transmission probability vector at an equilibrium.
\item[] $p_{\max}:$ upper bound to the transmission probability of a user.
\item[] $q_v:$ actual channel contention measure.
\item[] $q_v^*:$ theoretical channel contention measure.
\item[] $x^*:$ the limit of $K p$ as $K\to \infty$. The value of $x^*$ is obtained from the optimization of the asymptotic utility function.
\end{enumerate}

\section{Problem Formulation}
\label{SectionII}
Consider a wireless multiple access network with $K$ homogeneous users (transmitters) and a common receiver. The value of $K$ is known neither to the users nor to the receiver. Time is slotted such that each slot equals the length of one packet. Assume that each user has a saturated message queue. In each time slot, each individual user, say user $k$, makes its transmission/idling decision according to an associated transmission probability parameter, denoted by $p_k$. Transmission decision of a user is not shared with other users or with the receiver. We use transmission probability vector $\mbox{\boldmath $p$}=[p_{1}, \cdots, p_{K}]^T$ to denote the transmission probabilities of all users.

Let us add an integer-valued time index $t$ and use $\mbox{\boldmath $p$}(t)=[p_{1}(t), \cdots, p_{K}(t)]^T$ to denote the transmission probability vector of the users at the beginning of time slot $t$. Assume that users intend to maximize a symmetric network utility, which we will discuss later. In each time slot, according to channel feedback obtained from the receiver, we assume that user $k$ should derive a transmission probability target $\tilde{p}_k(t)$. User $k$ then updates its transmission probability by
\begin{equation}
p_k(t+1)=(1-\alpha(t))p_k(t) + \alpha(t) \tilde{p}_k (t),
\end{equation}
where $\alpha(t)\ge 0$ is the step size parameter of time slot $t$. Let $\mbox{\boldmath$\tilde{p}$}(t)=[\tilde{p}_{1}(t), \cdots, \tilde{p}_{K}(t)]^T$ be the vector of transmission probability targets of all users. Transmission probability vector $\mbox{\boldmath $p$}(t)$ is updated by
\begin{equation}
\mbox{\boldmath $p$}(t+1)=\mbox{\boldmath $p$}(t)+\alpha(t)(\mbox{\boldmath$\tilde{p}$}(t)-\mbox{\boldmath $p$}(t)).
\label{UpdatingAlgorithm}
\end{equation}
Probability adaptation given in (\ref{UpdatingAlgorithm}) falls into the stochastic approximation framework \cite{ref Kushner97}\cite{ref Karlin75}\cite{ref Borkar00}, and $\mbox{\boldmath$\tilde{p}$}(t)$ is often calculated based upon noisy estimates of certain system variables.

Let $\mbox{\boldmath$\hat{p}$}(t)=[\hat{p}_{1}(t), \cdots, \hat{p}_{K}(t)]^T$ be the vector of the theoretical transmission probability targets of all users, which is computed based upon presumed noiseless measurements and noiseless feedback in time slot $t$. Let $E_t[\mbox{\boldmath$\tilde{p}$}(t)]$ denote the conditional expectation of $\mbox{\boldmath$\tilde{p}$}(t)$ given system state at the beginning of time slot $t$. Let us write $E_t[\mbox{\boldmath$\tilde{p}$}(t)]$ as
\begin{equation}
E_t[\mbox{\boldmath$\tilde{p}$}(t)]=\mbox{\boldmath$\hat{p}$}(t)+\mbox{\boldmath $g$}(t)=\mbox{\boldmath$\hat{p}$}(\mbox{\boldmath $p$}(t))+\mbox{\boldmath $g$}(\mbox{\boldmath $p$}(t)),
\label{BiasTerm}
\end{equation}
where  $\mbox{\boldmath $g$}(t)$ is defined as the bias term in the probability vector target derivation. Given the communication channel, both $\mbox{\boldmath$\hat{p}$}(t)$ and $\mbox{\boldmath $g$}(t)$ are functions of $\mbox{\boldmath $p$}(t)$, which denotes the transmission probability vector in time slot $t$.

Next, we present two conditions that are typically required for the convergence of a stochastic approximation algorithm.
\begin{condition}
	\label{BiasCondition}
	(Mean and Bias) There exists a constant $K_m>0$ and a bounding sequence $0\le \beta(t) \le 1$, such that
	\begin{equation}
	\|\mbox{\boldmath$g$}(\mbox{\boldmath $p$}(t))\|\le K_m \beta(t).
	\end{equation}
    We assume that $\beta(t)$ is controllable in the sense that, for any $\epsilon>0$, one can design protocols to ensure $\beta(t)\le \epsilon$ for a sufficiently large $t$.
\end{condition}
\begin{condition}
	\label{lipschitz}	
	(Lipschitz Continuity) There exists a constant $K_l>0$, such that
	\begin{equation}
	\|\mbox{\boldmath$\hat{p}$}(\mbox{\boldmath $p$}_1)-\mbox{\boldmath$\hat{p}$}(\mbox{\boldmath $p$}_2)\|\le K_l\|\mbox{\boldmath $p$}_1-\mbox{\boldmath $p$}_2\|, \quad \mbox{for all } \mbox{\boldmath $p$}_1, \mbox{\boldmath $p$}_2.
	\end{equation}
\end{condition}

According to classical stochastic approximation theory \cite{ref Kushner97}\cite{ref Karlin75}\cite{ref Borkar00}, if Conditions \ref{BiasCondition} and \ref{lipschitz} are met, and $\alpha(t)$, $\beta(t)$ are small enough, trajectory of probability vector $\mbox{\boldmath $p$}(t)$ under distributed adaptation given in (\ref{UpdatingAlgorithm}) can be approximated by the following associated ODE
\begin{equation}
\frac{d\mbox{\boldmath $p$}(t)}{dt}=-\left[\mbox{\boldmath $p$}(t)-\mbox{\boldmath$\hat{p}$}(t)\right],
\label{ODE}
\end{equation}
where, with an abuse of the notation, we also used $t$ to denote the continuous time variable. Because all entries of $\mbox{\boldmath $p$}(t)$ and $\mbox{\boldmath$\hat{p}$}(t)$ stay in the range of $[0, 1]$, any equilibrium $\mbox{\boldmath $p$}^*$ of the associated ODE must satisfy $\mbox{\boldmath $p$}^*=\mbox{\boldmath$\hat{p}$}( \mbox{\boldmath $p$}^*)$.

Convergence of the distributed probability adaptation is stated in the following two theorems, which are quite standard for stochastic approximation algorithms. 

\begin{theorem}
	\label{strongconvergence}	
    Let Conditions \ref{BiasCondition} and \ref{lipschitz} hold. Assume that the associated ODE given in (\ref{ODE}) has a unique stable equilibrium at $\mbox{\boldmath$p$}^*$. If $\alpha(t)$ and $\beta(t)$ satisfy the following conditions
    \begin{equation}
    \sum_{t=0}^{\infty}\alpha(t)=\infty, \sum_{t=0}^{\infty}\alpha(t)^2<\infty, \sum_{t=0}^{\infty}\alpha(t)\beta(t)<\infty,
    \label{stepsize}
    \end{equation}
    then under distributed probability adaptation given in (\ref{UpdatingAlgorithm}), $\mbox{\boldmath$p$}(t)$ converges to $\mbox{\boldmath$p$}^*$ with probability one.
\end{theorem}

	Theorem \ref{strongconvergence} is implied by \cite[Theorems 4.3]{ref Kushner97}.

\begin{theorem}
	\label{weakconvergence}
    Let Conditions \ref{BiasCondition} and \ref{lipschitz} hold. Assume that the associated ODE given in (\ref{ODE}) has a unique stable equilibrium at $\mbox{\boldmath$p$}^*$. Under distributed probability adaptation given in (\ref{UpdatingAlgorithm}), for any $\epsilon>0$, there exists a constant $K_w>0$, such that, for any $0<\underline{\alpha}<\overline{\alpha}<1$ satisfying the following constraint
    \begin{equation}
    \exists T_0\ge 0, \underline{\alpha}\le \alpha(t) \le \overline{\alpha}, \beta(t)\le \sqrt{\overline{\alpha}}, \forall t\ge T_0,
    \label{bddstepsize}
    \end{equation}
    $\mbox{\boldmath$p$}(t)$ converges weakly to $\mbox{\boldmath$p$}^*$ in the following sense
    \begin{equation}
    \mathop{\lim\sup}_{t\to \infty} Pr\{\| \mbox{\boldmath$p$}(t)-\mbox{\boldmath$p$}^*\|\ge \epsilon \} < K_w\overline{\alpha}.
    \label{convergence}
    \end{equation}
\end{theorem}

Theorem \ref{weakconvergence} can be obtained by following the proof of \cite[Theorems 2.3]{ref Borkar00} with minor revisions.

Note that, in the above discussion, we assumed the same $\alpha(t)$ and $\beta(t)$ for all users. We also assumed that feedback information should be obtained by all users in every time slot, and probability adaptations of all users should be synchronous. However, by following the literature of stochastic approximation theory \cite{ref Kushner97}, it is easy to show that these assumptions can be relaxed. So long as step size sequences and bounding sequences of all users satisfy the same constraints given in (\ref{stepsize}) and (\ref{bddstepsize}), and all users receive channel feedback frequently enough, then conclusions of Theorems \ref{strongconvergence} and \ref{weakconvergence} should remain valid.

With convergence of the system guaranteed by Theorems \ref{strongconvergence} and \ref{weakconvergence}, key objectives of the system design are to develop the distributed MAC algorithm to satisfy Conditions \ref{BiasCondition} and \ref{lipschitz}, and to place the unique system equilibrium at a desired point that maximizes the chosen utility. Because users are homogeneous, due to symmetry, if a system equilibrium $\mbox{\boldmath$p$}^*$ is unique, it must take the form of $\mbox{\boldmath$p$}^*=p^* \mbox{\boldmath $1$}$, with $\mbox{\boldmath $1$}$ being the vector of all ones. That is, transmission probabilities of all users at the equilibrium must be identical. We choose to enforce such a property by requiring that all users should obtain the same transmission probability target in each time slot $t$, i.e., $\mbox{\boldmath$\hat{p}$}(t)=\hat{p}(t)\mbox{\boldmath$1$}$. The corresponding part of the system design is explained below.

Assume that there is a virtual packet being transmitted in each time slot. Virtual packets of different time slots are identical. A virtual packet is an assumed packet with coding parameters known both to the users and to the receiver, but it is not physically transmitted in the system, i.e., the packet is ``virtual''. We assume that, without knowing the transmission/idling status of the users, the receiver can detect in each time slot whether the virtual packet transmission should be regarded as successful or not. For example, suppose that the link layer channel is a collision channel, and a virtual packet has the same coding parameters of a real packet. Then, virtual packet reception in a time slot should be regarded as successful if and only if no real packet is transmitted. Success probability of the virtual packet in this case equals the idling probability of the collision channel. For another example, if all packets including the virtual packet are encoded using random block codes, given the physical layer channel, reception of each virtual packet corresponds to a detection task that judges whether or not the vector transmission status of all real users should belong to a specific region. Such detection tasks and their performance bounds have been extensively discussed in the distributed channel coding theory \cite{ref Luo12}\cite{ref Wang12}\cite{ref Luo15}.

Let $q_v(t)$ denote the success probability of the virtual packet in time slot $t$. We assume that the receiver should estimate $q_v(t)$ and feed it back to all transmitters. We term $q_v(t)$ the ``channel contention measure'' because it is designed to serve as a measurement of the contention level of the link-layer multiple access channel. Note that, in the collision channel case when $q_v(t)$ equals the channel idling probability, feeding back $q_v(t)$ may not be necessary. So long as each user $k$ knows the conditional success probability of its own packet, denoted by $q_k(t)$, idling probability of the channel can be estimated by $(1-p_k(t))q_k(t)$. With a general link layer channel, however, estimating $q_v(t)$ may not always be possible if it is not directly fed back from the receiver. Upon receiving the estimate of $q_v(t)$, each user calculates its probability target as the same function of the $q_v(t)$ estimate. Denote the theoretical transmission probability target by $\hat{p}(q_v(t))$. The theoretical vector transmission probability target is given by $\mbox{\boldmath$\hat{p}$}(t)= \hat{p}(q_v(t)) \mbox{\boldmath $1$}$. Because any equilibrium $\mbox{\boldmath $p$}^*$ of the ODE must satisfy $\mbox{\boldmath $p$}^*=\mbox{\boldmath$\hat{p}$}(\mbox{\boldmath $p$}^*)$, this guarantees that $\mbox{\boldmath $p$}^*$ must take the form of $\mbox{\boldmath $p$}^*=p^*\mbox{\boldmath$1$}$ with $p^*=\hat{p}(p^*)$, where $\hat{p}(p^*)$ is the theoretical transmission probability target computed under the assumption that all users should transmit with an identical probability of $p^*$.

In a practical system, the measurement of $q_v(t)$ is likely to be corrupted by noise. We assume that, if users keep their transmission probability vector $\mbox{\boldmath$p$}$ at a constant, and $q_v$ is measured over an interval of $Q$ time slots, then the measurement should converge to its true value with probability one as $Q$ is taken to infinity. Other than this assumption, measurement noise is not involved in the discussion of the design objectives of meeting Conditions \ref{BiasCondition} and \ref{lipschitz} and placing the unique system equilibrium at the desired point. Therefore, in the following section, we assume that $q_v(t)$ can be measured precisely and be fed back to the users. This leads to $\mbox{\boldmath $\tilde{p}$}(t)=\mbox{\boldmath $\hat{p}$}(t)= \hat{p}(t) \mbox{\boldmath $1$}$. We will also skip the time index $t$ to simplify the notations.

\section{Distributed MAC With Receiver Feeding Back The Channel Contention Measure}
\label{SectionIII}
In this section, we assume that success probability of the virtual packet can be measured at the receiver and be fed back to the transmitters. With a general link layer channel model, we will show that a distributed MAC algorithm can be developed to lead the transmission probabilities of all users to the same value that maximizes a chosen symmetric network utility.

We first introduce two sets of parameters to model the link layer channel. Define $\{C_{rj}\}$ for $j\ge 0$ as the ``real channel parameter set''. $C_{rj}$ is the conditional success probability of a real packet should it be transmitted in parallel with $j$ other real packets. Define $\{C_{vj}\}$ for $j\ge 0$ as the ``virtual channel parameter set''. $C_{vj}$ is the success probability of the virtual packet should it be transmitted in parallel with $j$ real packets. We assume that $C_{vj}\ge C_{v(j+1)}\ge 0$ should hold for all $j\ge 0$. This implies that an increased number of parallel real packet transmissions should not improve the chance of a virtual packet getting through the channel. Let $\epsilon_v\ge 0$ be a pre-determined small constant. Define $J_{\epsilon_v}$ as the minimum integer such that $C_{vJ_{\epsilon_v}}$ is strictly larger than $C_{v(J_{\epsilon_v}+1)}+\epsilon_v$, i.e.,
\begin{equation}
J_{\epsilon_v}=\mathop{\arg\min}_j C_{vj}>C_{v(j+1)}+\epsilon_v.
\label{J0Definition}
\end{equation}
Because $\{C_{rj}\}$ and $\{C_{vj}\}$ can be theoretically derived from the physical layer channel and coding parameters of the real and the virtual packets, we assume that they should be known to the users and also to the receiver. Note that, while $C_{rj}$ does not depend on the coding parameters of the virtual packet, coding design of the virtual packet does affect the value of $\{C_{vj}\}$.

We assume that the users intend to maximize a symmetric utility. Under the assumption that all users should transmit with the same probability, utility of the system is denoted by $U(K, p, \{C_{rj}\})$, which is defined as a function of the unknown number of users $K$, the common transmission probability $p$ of all users, and the real channel parameter set $\{C_{rj}\}$. For example, if users intend to maximize the sum throughput of the system, then $U(K, p, \{C_{rj}\})$ should be given by
\begin{eqnarray}
&& U(K, p, \{C_{rj}\}) \nonumber \\
&&=K\sum_{j=0}^{K-1} {K-1 \choose j} p^{j+1} (1-p)^{K-1-j} C_{rj}. \nonumber \\
\end{eqnarray}
For many utility functions of interest, such as the sum throughput function given above, an asymptotically optimal solution should maintain the expected load of the channel at a constant \cite{ref Hajek85}\cite{ref Ghez88}. Write $p=\frac{x}{K}$. We define $x^*$ using the following asymptotic utility optimization
\begin{equation}
x^*=\mathop{\mbox{argmax}}_x \lim \limits_{K\rightarrow\infty} U\left(K, \frac{x}{K}, \{C_{rj}\}\right).
\label{ValueOfXstar}
\end{equation}
The calculation of $x^*$ is only involved with the utility function and the real channel parameter set $\{C_{rj}\}$, and is irrelevant to the coding parameters of the virtual packet. We generally regard $p=\min\left\{1, \frac{x^*}{K}\right\}$ as an ideal solution to the utility optimization problem for all values of number of users. Note that, this is indeed the optimum solution for all $K$ for sum throughput maximization over a collision channel \cite{ref Hajek85}\cite{ref Ghez88}.

Let $b\ge 1$ be a pre-determined design parameter whose value will be introduced later. Define $p_{\max}$ as
\begin{equation}
p_{\max}=\min\left\{1, \frac{x^*}{J_{\epsilon_v}+b}\right\}.
\end{equation}
We will show next that, without knowing the actual number of users $K$, it is possible to set the unique system equilibrium at $\mbox{\boldmath $p$}^*=p^*\mbox{\boldmath$1$}=\min\{p_{\max}, \frac{x^*}{K+b}\}\mbox{\boldmath$1$}$, which is not far from the assumed ideal solution of $\min\{1, \frac{x^*}{K}\}\mbox{\boldmath$1$}$.

We intend to design a distributed MAC algorithm to set the unique system equilibrium at $p^*\mbox{\boldmath$1$}$ by maintaining channel contention at an appropriate level. Note that, given the virtual channel parameter set $\{C_{vj}\}$, channel contention measure $q_v(\mbox{\boldmath $p$}, K)$ is a function of the unknown number of users $K$ and the transmission probability vector $\mbox{\boldmath $p$}$. Because $q_v(\mbox{\boldmath $p$}, K)$ equals the summation of a finite number of polynomial terms, it should be Lipschitz continuous in $\mbox{\boldmath $p$}$ for any finite $K$. If all users transmit with the same probability $p$, i.e., $\mbox{\boldmath $p$}=p\mbox{\boldmath $1$}$, $q_v(p\mbox{\boldmath$1$}, K)$ is given by
\begin{equation}
	q_v(p\mbox{\boldmath$1$}, K)=\sum_{j=0}^K {K\choose{j}} p^j(1-p)^{K-j}C_{vj},
	\label{channelmodel}
\end{equation}
We assume that, upon receiving $q_v$ from the receiver, each user should first obtain an estimated number of users, denoted by $\hat{K}$, and then set the corresponding transmission probability target at $\tilde{p}=\hat{p}=\min \{p_{\max}, \frac{x^*}{\hat{K}+b}\}$, where $x^*>0$ is obtained from (\ref{ValueOfXstar}). We will show that, for any $x^*>0$, one can always find an appropriate $b$ and design a distributed MAC algorithm to ensure system convergence to the designed equilibrium of $\mbox{\boldmath$p$}^*=\min\{p_{\max}, \frac{x^*}{K+b}\}\mbox{\boldmath$1$}$. Note that, while the actual number of users $K$ is always an integer, we do allow the estimated number of users $\hat{K}$ to take a non-integer value.

Convergence of the MAC algorithm to be proposed depends on two key monotonicity properties presented below. First, the following theorem shows that, given the number of users $K$, $q_v(p\mbox{\boldmath$1$}, K)$ is non-increasing in $p$.
\begin{theorem}
	\label{PartialMonotonicity}
    With $C_{vj}\ge C_{v(j+1)}$ for all $j\ge 0$, $q_v(p\mbox{\boldmath$1$}, K)$ given in (\ref{channelmodel}) satisfies $\frac{\partial q_v(p\mbox{\boldmath$1$}, K)}{\partial p} \le 0$. Furthermore, $\frac{\partial q_v(p\mbox{\boldmath$1$}, K)}{\partial p} <0$ holds with strict inequality for $K>J_{\epsilon_v}$ and $p\in (0, 1)$.
\end{theorem}

The proof of Theorem \ref{PartialMonotonicity} is presented in \ref{ProofofPartialMonotonicity}.

Next, we define the ``theoretical channel contention measure'', denoted by $q^*_v$, which represents the expected channel contention level at the system equilibrium if the estimated number of users is correct. Let $\hat{p}=\frac{x^*}{\hat{K}+b}$, and $N=\lfloor\hat{K}\rfloor$ be the largest integer below $\hat{K}$. We define theoretical channel contention measure as a continuous function $q^*_v(\hat{p})$, which can also be viewed as a function of $\hat{K}$, as follows
\begin{equation}
q^*_v(\hat{p}) =\frac{\hat{p}-p_{N+1}}{p_{N}-p_{N+1}}q_N(\hat{p})+\frac{p_N-\hat{p}}{p_{N}-p_{N+1}}q_{N+1}(\hat{p}),
\label{qvstar}
\end{equation}
where
\begin{eqnarray}
&& p_N=\min\left\{p_{\max}, \frac{x^*}{N+b}\right\} \nonumber\\
&& p_{N+1}=\min\left\{p_{\max}, \frac{x^*}{N+1+b}\right\},
\end{eqnarray}
and
\begin{eqnarray}
&& q_N(p)=\sum_{j=0}^{N}{{N}\choose{j}} p^j(1-p)^{N-j}C_j \nonumber \\
&& q_{N+1}(p)=\sum_{j=0}^{N+1}{{N+1}\choose{j}} p^j(1-p)^{N+1-j}C_j. \nonumber \\
\label{successprob1}
\end{eqnarray}
If the number of users in the system indeed equals $K=\hat{K}$ with $\hat{K}\ge x^*-b$, then $q^*_v(\hat{p})$ defined in (\ref{qvstar}) equals the actual channel contention measure $q_v(\mbox{\boldmath $p$}^*, K)$ at the desired equilibrium $\mbox{\boldmath $p$}^*=\frac{x^*}{K+b}\mbox{\boldmath$1$}=\frac{x^*}{\hat{K}+b}\mbox{\boldmath$1$}$, i.e., when all users transmit with the same probability of $\hat{p}=\frac{x^*}{\hat{K}+b}$.

We intend to design the theoretical channel contention measure as a decreasing function in the estimated number of users $\hat{K}$. In other words, an increased number of users should lead to a more crowded channel. Equivalently, when being viewed as a function of $\hat{p}$, $q^*_v(\hat{p})$ is desired to be increasing in $\hat{p}$. Indeed, given an arbitrary $x^*>0$, such a monotonicity property can be guaranteed with an appropriate choice of $b$.
\begin{theorem}\label{monotonicity}
	Let $x^*>0$. If $b\ge \max\{1, x^*-\gamma_{\epsilon_v}\}$, with $\gamma_{\epsilon_v}$ being given by
\begin{eqnarray}
&& \gamma_{\epsilon_v}=\min_{N, N\ge J_{\epsilon_v}, N\ge x^*-b}  \nonumber \\
&& \frac{\sum_{j=0}^N j {N \choose j} \left(\frac{p_{N+1}}{1-p_{N+1}}\right)^j(C_{vj}-C_{v(j+1)}) }{\sum_{j=0}^N {N \choose j} \left(\frac{p_{N+1}}{1-p_{N+1}}\right)^j(C_{vj}-C_{v(j+1)})},
\label{GammaEpsilonvDefinition}
\end{eqnarray}
then $q_v^*(\hat{p})$ defined in (\ref{qvstar}) is non-decreasing in $\hat{p}$. Furthermore, if $b> \max\{1, x^*-\gamma_{\epsilon_v}\}$ holds with strict inequality, then $q_v^*(\hat{p})$ is strictly increasing in $\hat{p}$ for $\hat{p} \in (0, p_{\max})$.
\end{theorem}

The proof of Theorem \ref{monotonicity} is presented in \ref{Proofofmonotonicity}. We want to point out that, if $\epsilon_v$ is small enough to satisfy $C_{vj}=C_{v(j+1)}$ for all $j< J_{\epsilon_v}$, then we have $\gamma_{\epsilon_v}=J_{\epsilon_v}$. Otherwise, $\gamma_{\epsilon_v}\le J_{\epsilon_v}$ is generally true.

We are now ready to propose the distributed MAC algorithm.

{\bf Distributed MAC algorithm:}

\begin{enumerate}
	\item Initialize the transmission probabilities of all users. Let the transmission probability of user $k$ be denoted by $p_k$.
	\item Over an interval of $Q$ time slots, with $Q \ge 1$, the receiver measures the success probability of a virtual packet, denoted by $q_v$, and feeds $q_v$ back to all transmitters. \label{repeat}
	\item Upon receiving $q_v$, each user (transmitter) derives a transmission probability target $\hat{p}$ by solving the following equation
	\begin{equation}
	q_v^*(\hat{p})=q_v.
	\label{updatecriterion}
	\end{equation}
	If a $\hat{p}\in \left[0, p_{\max}\right]$ satisfying (\ref{updatecriterion}) cannot be found, each user sets $\hat{p}$ at $\hat{p}=p_{\max}$ when $q_v > q_v^*(p_{\max})$, or at $\hat{p}=0$ when $q_v < q_v^*(0)$.
	\item User $k$ then updates its transmission probability by
	\begin{equation}
	p_k=(1-\alpha)p_k+\alpha\hat{p},
	\end{equation}
	where $\alpha$ is the step size parameter for user $k$.
	\item The process is repeated from Step \ref{repeat} till transmission probabilities of all users converge.
\end{enumerate}

Convergence of the proposed MAC algorithm is stated in the following theorem.
\begin{theorem}{\label{MACConvergence}}
	Given $x^*>0$, let $b$ be chosen to satisfy $b>\max\{1, x^*-\gamma_{\epsilon_v}\}$. With the proposed MAC algorithm, the system has a unique equilibrium at $\mbox{\boldmath$p$}^*=\min\{p_{\max}, \frac{x^*}{K+b}\}\mbox{\boldmath$1$}$. Furthermore, given the number of users $K$, the probability target $\hat{p}( \mbox{\boldmath$p$})$ as a function of transmission probability vector $\mbox{\boldmath$p$}$ satisfies Conditions \ref{BiasCondition} and \ref{lipschitz}. Consequently, transmission probability vector $\mbox{\boldmath$p$}$ converges to $\mbox{\boldmath$p$}^*$ in the sense explained in Theorems \ref{strongconvergence} and \ref{weakconvergence}.
\end{theorem}

The proof of Theorem \ref{MACConvergence} is presented in \ref{ProofofMACConvergence}.

The above analysis indicates that, with an arbitrary virtual packet design and with the proposed MAC algorithm, the system should converge to the designed equilibrium so long as $b$ is chosen to satisfy $b>\max\{1, x^*-\gamma_{\epsilon_v}\}$. However, one should note that optimality of the algorithm does depend on the value of $b$, $\gamma_{\epsilon_v}$, and $J_{\epsilon_v}$, which are determined by the virtual channel parameter set $\{C_{vj}\}$, and therefore are dependent on the coding parameters of the virtual packet. For example, it is known that, to maximize the sum throughput of a distributed multiple access system over a collision channel, the optimal solution is to set the transmission probability of all users at $p=\frac{1}{K}$ with $K$ being the number of users. This corresponds to $x^*=1$ and $b=0$ in our model. For a general system, assume that setting the transmission probabilities of all users at $p=\min\{1, \frac{x^*}{K}\}$ should be an ideal choice for maximizing the chosen utility. Because the proposed MAC algorithm sets the equilibrium at $\mbox{\boldmath$p$}^*=\min\{p_{\max}, \frac{x^*}{K+b}\}\mbox{\boldmath$1$}$, there are two optimality concerns. On one hand, when the number of users $K$ is large, it is a general preference that one should design the virtual packet to allow a relatively small value of $b$. This implies that the values of $\gamma_{\epsilon_v}$ and $J_{\epsilon_v}$ should not be much smaller than $x^*$. On the other hand, when the actual number of users $K$ is small, one should also try to get $p_{\max}$ close to $1$. This implies that $J_{\epsilon_v}$ should also not be much larger than $x^*$. Combining both optimality concerns, a general guideline is to design coding parameters of the virtual packet such that $J_{\epsilon_v}$ and $\gamma_{\epsilon_v}$ should be slightly smaller than $x^*$ and $b$ should be close to $1$.

\section{Distributed MAC with Interpreted Channel Contention Measure}
\label{SectionIV}
Classical MAC protocols often assume that a user should get feedback from the receiver on whether its own packets are successfully received or not \cite{ref Bertsekas92}. This enables each user, say user $k$, to measure the conditional success probability of its own packet transmissions, denoted by $q_k$. In this section, we consider the case when $q_k$ is the only feedback available to user $k$. To simplify the discussion, we also assume that a virtual packet should have the same communication parameters as those of a real packet. In order to apply the MAC algorithm proposed in Section \ref{SectionIII}, user $k$ will need to interpret the success probability of the virtual packet based on the measurement of $q_k$. Because transmission activities of the users are mutually independent, under the assumption that a virtual packet should have the same coding parameters of a real packet, $q_k$ equals the conditional success probability of the virtual packet given that user $k$ idles. Consequently, user $k$ can calculate the success probability of the virtual packet by
\begin{equation}
q_v=(1-p_k)q_k+p_k d_k,
\label{QvDecomposition}
\end{equation}
where $p_k$ is the transmission probability of user $k$, and $d_k$ is the conditional success probability of the virtual packet given that user $k$ transmits a packet\footnote{Extensions can be made to the case when a virtual packet is equivalent to the combination of $R$ real packets by decomposing $q_k$ in a similar way as shown in (\ref{QvDecomposition}).}. Note that $d_k$ can be easily calculated in special cases. For example, under a collision channel model, we have $d_k=0$. In this case, $q_v=(1-p_k)q_k$ is the actual success probability of the virtual packet. However, for a general channel, $d_k$ may not always be available at the transmitters unless additional feedback information is provided. When $d_k$ is not available, we propose a two-step approach for each user to interpret $d_k$ and hence the success probability of the virtual packet $q_v$, and then to update its transmission probability accordingly.

To explain the detail of the two-step approach, we need to define two auxiliary functions. More specifically, for an arbitrary estimated number of users $\breve{K}$, let $\breve{N}=\lfloor \breve{K}\rfloor$ denote the largest integer below $\breve{K}$. Let $\breve{p}=\min\{p_{\max}, \frac{x^*}{\breve{K}+b}\}$, $p_{\breve{N}}=\min\{p_{\max}, \frac{x^*}{\breve{N}+b}\}$ and $p_{\breve{N}+1}=\min\{p_{\max}, \frac{x^*}{\breve{N}+1+b}\}$, where $b$ is a constant satisfying $b>\max\{1, x^*-\gamma_{\epsilon_v}\}$. We define auxiliary functions $q^*(\breve{p})$ and $d^*(\breve{p})$ as follows
\begin{eqnarray}
&&q^*(\breve{p})= \frac{\breve{p}-p_{\breve{N}+1}}{p_{\breve{N}}-p_{\breve{N}+1}}\sum_{j=0}^{\breve{N}-1} {{\breve{N}-1}\choose{j}}\nonumber \\
&& \qquad \times \breve{p}^j(1-\breve{p})^{\breve{N}-1-j} C_{vj}\nonumber \\
&& \quad +\frac{p_{\breve{N}}-\breve{p}}{p_{\breve{N}}-p_{\breve{N}+1}} \sum_{j=0}^{\breve{N}} {{\breve{N}}\choose{j}} \breve{p}^j(1-\breve{p})^{\breve{N}-j}C_{vj}, \nonumber \\
&&d^*(\breve{p})= \frac{\breve{p}-p_{\breve{N}+1}}{p_{\breve{N}}-p_{\breve{N}+1}}\sum_{j=0}^{\breve{N}-1} {{\breve{N}-1}\choose{j}} \nonumber \\
&& \qquad \times \breve{p}^j(1-\breve{p})^{\breve{N}-1-j}C_{v(j+1)} \nonumber \\
&& \quad +\frac{p_{\breve{N}}-\breve{p}}{p_{\breve{N}}-p_{\breve{N}+1}}\sum_{j=0}^{\breve{N}} {{\breve{N}}\choose{j}} \breve{p}^j(1-\breve{p})^{\breve{N}-j}C_{v(j+1)}. \nonumber \\
\label{qstar}
\end{eqnarray}
In the case when $\breve{K}$ takes an integer value, $q^*(\breve{p})$ is the conditional success probability of the virtual packet under the assumptions that the system has $\breve{K}$ users, user $k$ idles, and all other users have the same transmission probability of $\breve{p}$. Similarly, $d^*(\breve{p})$ represents the conditional success probability of the virtual packet under the assumptions that the system has $\breve{K}$ users, user $k$ transmits a packet, and all other users have the same transmission probability of $\breve{p}$.

Next, we present the two-step approach that is suggested for each user to obtain its transmission probability target.

{\bf Step 1:} $\quad$ Over an interval of $Q\ge 1$ time slots, each user, say user $k$, measures its own conditional success probability $q_k$. User $k$ then obtains an intermediate transmission probability $\breve{p}$ by solving the following equation
\begin{equation}
q^*(\breve{p})=q_k.
\label{FirstStepTarget}
\end{equation}
If a $\breve{p} \in \left[0, p_{\max}\right]$ satisfying (\ref{updatecriterion}) cannot be found, user $k$ sets $\breve{p}$ at $\breve{p}=p_{\max}$ when $q_k > q^*(p_{\max})$, or at $\breve{p}=0$ when $q_k < q^*(0)$ .

{\bf Step 2:} $\quad$ In the second step, user $k$ interprets channel contention measure $q_v$ as
\begin{equation}
q_v=(1-p_k)q_k+p_k d^*(\breve{p}).
\end{equation}
An updated transmission probability target $\hat{p}$ for user $k$ is then determined by solving equation (\ref{updatecriterion}). As before, if a $\hat{p}\in \left[0, p_{\max}\right]$ satisfying (\ref{updatecriterion}) cannot be found, user $k$ sets $\hat{p}$ at $\hat{p}=p_{\max}$ when $q_v > q_v^*(p_{\max})$, or at $\hat{p}=0$ when $q_v < q_v^*(0)$.

Note that when $\hat{p}$ is obtained by the two step approach, a convergence proof of the MAC algorithm is no longer available. This is because the two step approach does not guarantee that transmission probability targets obtained by different users should be identical. Therefore, the assumption that any equilibrium $\mbox{\boldmath$p$}^*$ must take the form of $\mbox{\boldmath$p$}^*=p^*\mbox{\boldmath$1$}$ is no longer valid. Nevertheless, in the following theorem, we show that the two-step approach is equivalent to a simplified one-step approach where user $k$ directly uses $\breve{p}$ obtained in (\ref{FirstStepTarget}) as its transmission probability target.
\begin{theorem} \label{onestep}
Let $x^*>0$, and $b\ge \max\{1, x^*-\gamma_{\epsilon_v}\}$, where $\gamma_{\epsilon_v}$ is defined in (\ref{GammaEpsilonvDefinition}). Suppose that each user, say user $k$, first obtains an intermediate transmission probability $\breve{p}$ and then determines its transmission probability target $\hat{p}$ by following the two-step approach. Then $\breve{p}\ge p_k$ implies $\hat{p}\ge p_k$, while $\breve{p}\le p_k$ implies $\hat{p}\le p_k$.
\end{theorem}

The proof of Theorem \ref{onestep} is presented in \ref{Proofofonestep}.

Theorem \ref{onestep} suggests that each user can simplify the two step approach into Step 1 only and simply set the transmission probability target at $\hat{p}=\breve{p}$. In cases when the two-step approach does lead the system to the designed equilibrium, the simplified one step approach should also lead the system to the same equilibrium.

\section{Simulation Results}
\label{SectionV}
In this section, we use computer simulations to illustrate both the optimality and the convergence properties of the proposed MAC algorithms.

{\bf Example 1:} (Optimality) The distributed MAC framework proposed in this paper shares certain proximity with the one proposed in \cite{ref Hajek85}, although \cite{ref Hajek85} only considered the simple collision channel model. In the first example, we investigate the classical problem of symmetric sum throughput maximization over a collision channel.

Assume that the system has $K$ users each having a saturated message queue. If $K$ is known, the optimal solution that maximizes the sum throughput is to set the transmission probabilities of all users at $p_{\mbox{\scriptsize opt}}=\frac{1}{K}$ \cite{ref Hajek85}. It was suggested in \cite{ref Hajek85} that, when $K$ is unknown, with the help of a proposed distributed MAC framework, each user should direct its transmission probability to converge to $p_a$, which is obtained by solving the following equation.
\begin{equation}
e P(\mbox{idle}) -1 - 0.5 \sqrt{p_a} =0, \quad P(\mbox{idle})=(1-p_a)^K,
\end{equation}
where $P(\mbox{idle})$ is the idling probability of the channel that can be measured locally if each user knows the conditional success probability of its own packets.

Now consider the distributed MAC algorithm proposed in Section \ref{SectionIII} of this paper. With the collision channel model, the real channel parameter set $\{C_{rj}\}$ is given by $C_{r0}=1$ and $C_{rj}=0$ for all $j>0$. Because the utility function is chosen as the sum network throughput, we obtain from (\ref{ValueOfXstar}) that $x^*=1$. Let us assume that a virtual packet should have the same coding parameters as those of a real packet. Consequently, the virtual channel parameter set $\{C_{vj}\}$ is identical to the real channel parameter set, i.e., $C_{vj}=C_{rj}$ for all $j\ge 0$. Choose $\epsilon_v=0.01$, we get $\gamma_{\epsilon_v}=J_{\epsilon_v}=0$. This supports the choice of $b=1.01 > x^*-\gamma_{\epsilon_v}$. The unique equilibrium of the system is therefore set at $\mbox{\boldmath$p$}^*=p^*\mbox{\boldmath$1$}=\frac{1}{K+1.01}\mbox{\boldmath$1$}$.

\begin{figure}[ht]
\begin{center}
    \includegraphics[width=3 in]{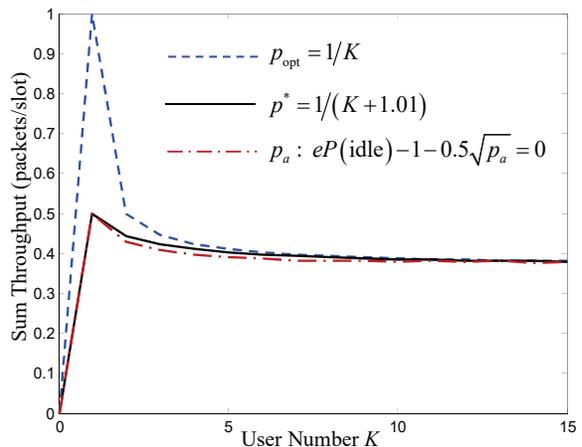}\caption{\label{Figure1}  Sum throughput as functions of the number of users for distributed multiple access over a collision channel.}
\end{center}
\end{figure}
In Figure \ref{Figure1}, we illustrate the achieved sum throughput in packets/slot as functions of the number of users under various system settings. The dashed curve represents the optimum utility when transmission probabilities of all users are set at $p_{\mbox{\scriptsize opt}}$. Note that the optimum utility is not necessarily achievable because it requires the knowledge of the number of users $K$. The solid curve represents the achieved utility at the designed equilibrium of the proposed MAC algorithm, i.e., when transmission probabilities of all users are set at $p^*$. The dash-dotted curve is the achieved utility when transmission probabilities of all users are set at $p_a$, as suggested in \cite{ref Hajek85}. It can be seen that, at the designed equilibrium, the proposed distributed MAC algorithm can achieve a throughput performance slightly higher than the approach suggested in \cite{ref Hajek85}.

{\bf Example 2:} (Optimality) In this example, we consider a distributed multiple access network with $K$ users and a simple fading channel. In each time slot, with a probability of $0.3$, the channel can support no more than $M_1=4$ parallel real packet transmissions, and with a probability of $0.7$, the channel can support no more than $M_2=6$ parallel real packet transmissions. Note that such a channel can appear if there is an interfering user that transmits a packet with probability $0.3$ in each time slot. One packet from the interfering user is equivalent to the combination of two packets from a regular user. The real channel parameter set $\{C_{rj}\}$ in this case is given by $C_{rj}=1$ for $j<4$, $C_{rj}=0.7$ for $4\le j<6$, and $C_{rj}=0$ for $j\ge 6$. Assume that users intend to optimize the symmetric throughput weighted by a transmission energy cost of $E=0.3$. With the number of users being $K$ and all users transmitting with the same probability $p$, system utility $U(K, p, \{C_{rj}\})$ is given by
\begin{eqnarray}
\label{utility1}
&& U(K, p, \{C_{rj}\})= - E Kp + \nonumber \\
&& \quad \sum_{j=0}^{K-1}K{K-1 \choose j} p^{j+1} (1-p)^{K-1-j}C_{rj}.
\end{eqnarray}
Correspondingly, $x^*$ can be obtained from (\ref{ValueOfXstar}) as $x^*=3.29$. Assume that a virtual packet should have the same coding parameters as those of a real packet. The virtual channel parameter set $\{C_{vj}\}$ is therefore identical to the real channel parameter set, i.e., $C_{vj}=C_{rj}$ for all $j\ge 0$. With $\epsilon_v = 0.01$, we have $\gamma_{\epsilon_v}=J_{\epsilon_v} = 3$. Therefore, we can set $b = 1.01> x^*-\gamma_{\epsilon_v}$.

\begin{figure}[!ht]
\begin{center}
    \includegraphics[width=3 in]{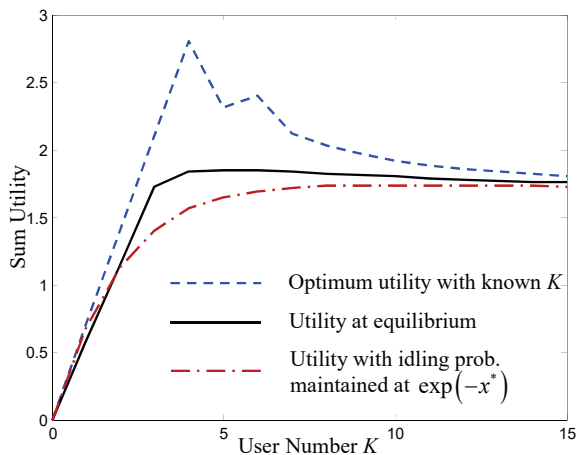}\caption{\label{Figure2}  Sum utility as functions of the number of users for distributed multiple access over a simple fading channel.}
\par\end{center}
\end{figure}

In Figure \ref{Figure2}, we illustrate three utilities all as functions of the number of users $K$. The solid curve represents the utility achieved by the proposed MAC algorithm at the designed equilibrium. The dashed curve represents the optimum utility under the assumption that number of users $K$ is known, and this is not necessarily achievable without the knowledge of $K$. The dash-dotted curve represents the utility if we maintain the channel idling probability at its asymptotically optimal value of $\exp(-x^*)$, as suggested in \cite{ref Hajek85}\footnote{While \cite{ref Hajek85} also suggested to maintain other variables at their asymptotically optimal values, these alternative approaches do not lead to a better performance in this example.}. This is equivalent to setting the transmission probabilities of all users at $1-\exp\left(-\frac{x^*}{K}\right)$. It can be seen that, the proposed MAC algorithm can achieve a higher utility value compared with the approach suggested in \cite{ref Hajek85}. Achieved utility of the proposed MAC algorithm is also reasonably close to optimal when the number of users $K$ is not close to $M$.

{\bf Example 3:} (Convergence with receiver feeding back the channel contention measure) Following Example 2, we have $x^*=3.29$ and $b=1.01$. Assume that the system has $K=8$ users. We initialize the transmission probabilities of all users at $0$. In each time slot, a channel state flag is randomly generated to indicate whether the channel can support the parallel transmissions of no more than $4$ or $6$ packets. Each user also randomly determines whether a packet should be transmitted according to its own transmission probability parameter. Whether the real packets and the virtual packet can go through the channel or not is then determined using the corresponding channel model. We use the following exponential moving average approach to measure $q_v$. $q_v$ is initialized at $q_v=1$. In each time slot, $q_v$ is updated by $q_v=(1-\frac{1}{300})q_v+\frac{1}{300}I_v$, where $I_v\in \{0, 1\}$ is an indicator of the success/failure status of the virtual packet in the current time slot, i.e., $I_v=1$ indicates that the transmission of the virtual packet in this time slot should be regarded as successful and $I_v=0$ otherwise. While this is different from the approach proposed in the distributed MAC algorithm, simulations show that an exponential averaging measurement of $q_v$ can often lead the system to converge in a relatively smaller number of time slots. We assume that $q_v$ is measured at the receiver and is then fed back in each time slot to all transmitters. The rest of probability adaptation proceeds according to the distributed MAC algorithm introduced in Section \ref{SectionIII} with a constant step size of $\alpha=0.05$.

\begin{figure}[!ht]
\begin{center}
	\includegraphics[width=3 in]{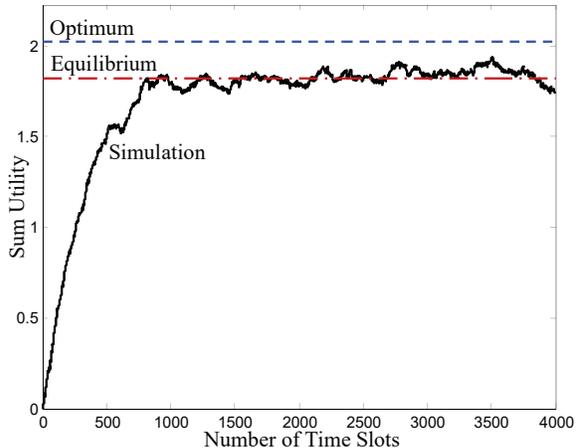}\caption{\label{Figure3}  Convergence in system utility of a multiple access network with $K=8$ users over a simple fading channel. Channel contention measure is fed back by the receiver.}
\par\end{center}
\end{figure}
Convergence behavior of the system utility is illustrated in Figure \ref{Figure3}, where system utility is also measured using the same exponential moving average approach except that initial value of the utility is set at $0$. The dash-dotted line represents the system utility if each user transmits with the desired probability $p^*=\frac{x^*}{K+b}=0.365$. The dashed line represents the optimal system utility obtained by $\max_{0\le p \le 1} U(K, p, \{C_{rj}\})$, where $U(K, p, \{C_{rj}\})$ is given in (\ref{utility1}). In this case, system utility at the designed equilibrium is about $90\%$ of the optimal value. In about $1000$ interations, transmission probabilities of all users already become close to the equilibrium value.

{\bf Example 4:} (Convergence with interpreted channel contention measure) In this example, we study the convergence property of the distributed MAC algorithm proposed in Section \ref{SectionIV} when each user only knows the success/failure status of its own packets. Following Example 3, we assume that each user, say user $k$, should maintain a measurement of the conditional success probability of its own packets, denoted by $q_k$. $q_k$ is initialized at $q_k=1$. In each time slot, if user $k$ transmits a packet, then $q_k$ is updated by $q_k=(1-\frac{1}{300})q_k+\frac{1}{300}I_k$, where $I_k\in \{0, 1\}$ is an indicator of the success/failure status of the packet transmitted by user $k$ in the current time slot. We assume that the value of $I_k$ should be fed back to user $k$ from the receiver. If user $k$ idles, on the other hand, the value of $q_k$ should remain unchanged. With the measurement of $q_k$, user $k$ then uses the simplified one step approach to derive its transmission probability target $\hat{p}=\breve{p}$ by solving equation (\ref{FirstStepTarget}). Then, user $k$ updates its transmission probability by $p_k=(1-\alpha)p_k+\alpha \hat{p}$ with a constant step size of $\alpha=0.05$.

\begin{figure}[!ht]
\begin{center}
	\includegraphics[width=3 in]{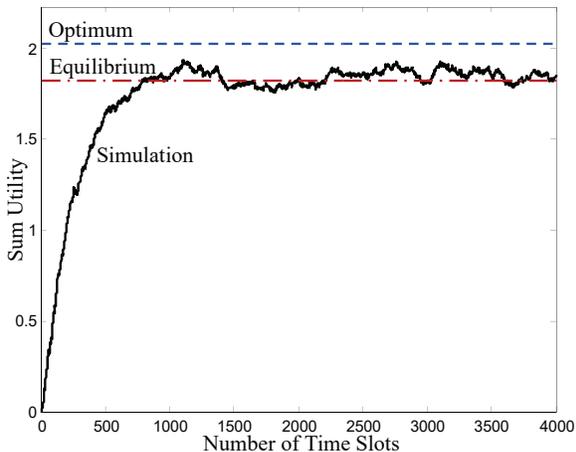}\caption{\label{Figure4}  Convergence in system utility of a multiple access network with $K=8$ users over a fading channel. Each user only knows the success/failure status of its own packets.}
\par\end{center}
\end{figure}

In Figure \ref{Figure4}, we illustrate the convergence behavior in sum utility of the system. As before, system utility is measured using the same exponential moving average approach with an initial value of $0$. It can be seen that, when each user uses the one step approach to calculate its transmission probability target, the system can still converge to the same designed equilibrium.

{\bf Example 5:} (Convergence in a dynamic environment) In this example, we start with the system introduced in Example 4. We still assume that each user only knows the success/failure status of its own packets, and uses the simplified one step approach to calculate its transmission probability target. The system contains $K=8$ users at the beginning. We say that the system starts with Stage 1. At the $3001$th time slot, we assume that the system enters Stage 2 when $7$ other users join the network. This leads to a total of $K=15$ users. Each of the new users has its transmission probability initialized at zero and its packet conditional success probability initialized at one. Then at the $6001$th time slot, we assume that the system enters Stage 3 when $5$ users exit the network.

\begin{figure}[!ht]
\begin{center}
	\includegraphics[width=3 in]{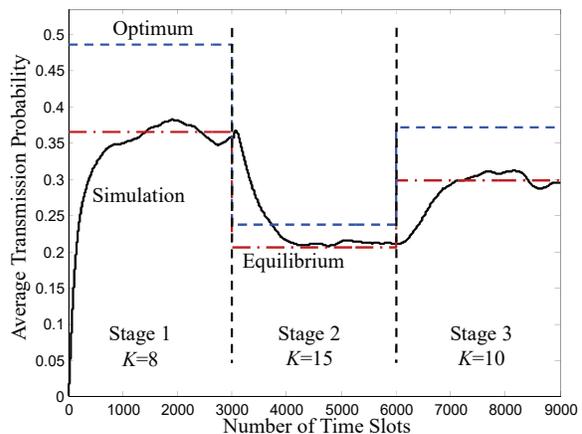}\caption{\label{Figure5}  Convergence in average transmission probability of the active users of a multiple access network over three stages.}
\end{center}
\end{figure}

In Figure \ref{Figure5}, we illustrate convergence behavior of the system in average transmission probability of the active users over the three stages. The corresponding optimal transmission probability (i.e., transmission probability that maximizes the symmetric utility) and the theoretical transmission probability at the designed equilibria are also illustrated in dashed lines and dash-dotted lines, respectively. While a theoretical convergence proof is not available in this case, we can see that in a dynamic environment when users join/exit the system, the proposed MAC algorithm has a reasonably good capability to help active users tracking the designed equilibrium.

\section{Conclusion}
\label{SectionVI}
We investigated multiple access networking with an unknown finite number of homogeneous users. A distributed MAC algorithm is proposed to maximize an arbitrarily chosen symmetric network utility with a generally modeled link layer channel. We proposed to measure contention level of the channel using the success probability of a carefully designed virtual packet, and to adapt transmission probabilities of all users toward a direction that matches the actual channel contention measure with its theoretical value. Under the assumption that channel contention measure can be fed back by the receiver, we proved the convergence of the proposed MAC algorithm with the help of two key monotonicity properties. We also proposed a revised MAC algorithm for the case when each user only knows the success/failure status of its own packets. While a convergence proof of the revised MAC algorithm is not available, simulation results suggest that the revised MAC algorithm can indeed lead the system to the same designed equilibrium.

\appendix
\section{Proof of Theorem \ref{PartialMonotonicity}}
\label{ProofofPartialMonotonicity}
	Partial derivative of $q_v(p\mbox{\boldmath$1$}, K)$ with respect to $p$ is given by
	\begin{eqnarray}
	&& \frac{\partial q_v(p\mbox{\boldmath$1$}, K)}{\partial p}=\sum_{j=0}^{K} {{K}\choose{j}} j p^{j-1}(1-p)^{K-j}C_{vj} \nonumber \\
	&& \quad -\sum_{j=0}^{K} {{K}\choose{j}} p^{j} (K-j) (1-p)^{K-j-1}C_{vj} \nonumber \\
	&& = -\sum_{j=0}^{K-1} K{{K-1}\choose{j}}p^{j}(1-p)^{K-1-j} \nonumber \\
    && \quad \times(C_{vj}-C_{v(j+1)}) \nonumber \\
    && \le 0,
    \label{QvDerivative}
	\end{eqnarray}
	where the last inequality is due to the assumption that $C_{vj}\ge C_{v(j+1)}$ for all $j\ge 0$. (\ref{QvDerivative}) holds with strict inequality if $K>J_{\epsilon_v}$ and $p(1-p)\ne 0$.

\section{Proof of Theorem \ref{monotonicity}}
\label{Proofofmonotonicity}

Let us first consider the situation when $\frac{x^*}{N+b}\le p_{\max}$. According to the definition of  $q^*_v(\hat{p})$ given in (\ref{qvstar}), we have
\begin{eqnarray}
	&&\frac {dq^*_v(\hat{p})}{d\hat{p}} =\frac {q_N(\hat{p})-q_{N+1}(\hat{p})}{p_{N}-p_{N+1}} \nonumber \\
    && \quad + \frac{\hat{p}-p_{N+1}}{p_{N}-p_{N+1}}\frac {dq_N(\hat{p})}{d\hat{p}} + \frac{p_N-\hat{p}}{p_{N}-p_{N+1}}\frac {dq_{N+1}(\hat{p})}{d\hat{p}}. \nonumber \\
	\label{qderivative}
\end{eqnarray}
Write $\hat{K}=N+1-\lambda$ with $\lambda\in \left(0, 1\right]$. We have
\begin{equation}
	\hat{p}-p_{N+1}=\frac{x^*}{\hat{K}+b}-\frac{x^*}{N+1+b}=\frac{\lambda}{N+1+b}\hat{p},
	\label{plambda1}
\end{equation}
and
\begin{equation}
	p_N-\hat{p}=\frac{x^*}{N+b}-\frac{x^*}{\hat{K}+b}=\frac{1-\lambda}{N+b}\hat{p}.
	\label{plambda2}
\end{equation}	
Meanwhile, because $q_{N+1}(\hat{p})$ can be decomposed as
\begin{eqnarray}
	&& q_{N+1}(\hat{p})=\sum_{j=0}^{N+1} {{N+1}\choose{j}} \hat{p}^j(1-\hat{p})^{N+1-j}C_{vj} \nonumber \\
	&& =\hat{p}\sum_{j=0}^{N} {{N}\choose{j}} \hat{p}^j(1-\hat{p})^{N-j}C_{v(j+1)} \nonumber \\
	&& \quad +(1-\hat{p})\sum_{j=0}^{N} {{N}\choose{j}} \hat{p}^j(1-\hat{p})^{N-j}C_{vj},
	\label{qdecomposition}
\end{eqnarray}
we have
\begin{equation}
	q_N-q_{N+1}=\sum_{j=0}^{N} {{N}\choose{j}} \hat{p}^{j+1}(1-\hat{p})^{N-j}(C_{vj}-C_{v(j+1)}).
	\label{qdifference}
\end{equation}
Furthermore, by taking derivatives of $q_N(\hat{p})$ and $q_{N+1}(\hat{p})$ with respect to $\hat{p}$, we get,
\begin{eqnarray}
	&&\frac{dq_N(\hat{p})}{d\hat{p}} = \sum_{j=0}^{N} (N-j){{N}\choose{j}} \hat{p}^{j}(1-\hat{p})^{N-j-1} \nonumber \\
    && \qquad \times (C_{v(j+1)}-C_{vj}),
	\label{qNderivative}
\end{eqnarray}
and
\begin{eqnarray}
	&& \frac{dq_{N+1}(\hat{p})}{d\hat{p}}= \sum_{j=0}^{N} (N+1){{N}\choose{j}}\hat{p}^{j}(1-\hat{p})^{N-j} \nonumber \\
    && \qquad \times(C_{v(j+1)}-C_{vj}).
	\label{qNplus1derivative}
\end{eqnarray}	
Substituting the above results into (\ref{qderivative}) yields
\begin{eqnarray}
	&&(p_{N}-p_{N+1})\frac {dq^*_v(\hat{p})}{d\hat{p}}\nonumber\\
    &&= \sum_{j=0}^{N} \left( {N \atop j} \right) \hat{p}^{j+1} (1-\hat{p})^{N-j} \left( C_{vj}-C_{v(j+1)} \right) \nonumber \\
	&&  \quad -\frac{\lambda}{N+1+b} \sum_{j=0}^{N} (N-j) \left( {N \atop j} \right)   \nonumber \\
    && \qquad \times \hat{p}^{j+1}  (1-\hat{p})^{N-j-1}\left( C_{vj}-C_{v(j+1)} \right) \nonumber \\
	&& \quad - \frac{1-\lambda}{N+b} \sum_{j=0}^{N} (N+1) \left( {N \atop j} \right)  \nonumber \\
    && \qquad \times \hat{p}^{j+1} (1-\hat{p})^{N-j}\left( C_{vj}-C_{v(j+1)} \right) \nonumber \\
	&&=\sum_{j=0}^{N}{{N}\choose{j}}\hat{p}^{j+1}(1-\hat{p})^{N-j-1}(C_{vj}-C_{v(j+1)}) \nonumber\\
	&& \quad \times\left(\frac{\lambda((1-\hat{p})(N+1+b)-N+j)}{N+1+b} \right.\nonumber\\
	&& \quad \left. +\frac{(1-\lambda)(1-\hat{p})(b-1)}{N+b}\right).
    \label{Substitution}
\end{eqnarray}	

Note that, for all $j\ge 0$, we have
\begin{equation}
	\frac{\lambda((1-\hat{p})(N+1+b)-N+j)}{N+1+b}\ge \frac{\lambda(b-x^*+j)}{N+1+b}.
	\label{derivative}
\end{equation}
Therefore, $\frac{dq^*_v(\hat{p})}{d\hat{p}}\ge0$ if $b\ge 1$ and
\begin{equation}
\sum_{j=0}^{N}{{N}\choose{j}}\left(\frac{\hat{p}}{1-\hat{p}}\right)^{j}(C_{vj}-C_{v(j+1)})(b-x^*+j)\ge 0.
\label{GammaEpsilonvInequality}
\end{equation}
(\ref{GammaEpsilonvInequality}) holds if $b\ge x^*-\gamma_{\epsilon_v}$ with $\gamma_{\epsilon_v}$ being defined in (\ref{GammaEpsilonvDefinition}).

Furthermore, if we have both $b>1$ and $b>x^*-\gamma_{\epsilon_v}$ holding with strict inequalities, and $N\ge J_{\epsilon_v}$, then $\frac{dq^*_v(\hat{p})}{d\hat{p}}>0$ should also hold with strict inequality for $\hat{p} \in (0,p_{max})$.

Now consider the situation when $\frac{x^*}{N+b} \ge p_{max}$. It is easy to see that, when $\frac{x^*}{\hat{K}+b} \ge p_{max}$, we have $\frac{d q_v^*(\hat{p})}{d \hat{p}}= 0$. When $\frac{x^*}{\hat{K}+b} < p_{max}$ but $\frac{x^*}{N+b} \ge p_{max}$, on the other hand, we can write $\hat{K}=N+1-\lambda$ with $0 < \lambda \le N+1+b-\frac{x^*}{p_{max}}$. Consequently, (\ref{qderivative}) and (\ref{plambda1}) still hold, but (\ref{plambda2}) should be replaced by
\begin{equation}
p_{N}-\hat{p}=p_{max} - \frac{x^*}{\hat{K}+b} \leq \frac{1-\lambda}{N+b} \hat{p}.
\label{plambda2modified}
\end{equation}
Therefore, (\ref{Substitution}) becomes
\begin{eqnarray}
&& \left( p_N-p_{N+1} \right) \frac{d q_v^*\left( \hat{p} \right)}{d \hat{p}} \nonumber \\
&& \ge \sum_{j=0}^{N} \left( {N \atop j} \right) \hat{p}^{j+1} (1-\hat{p})^{N-j-1} \left( C_{vj}-C_{v(j+1)} \right) \nonumber \\
&& \quad \times \left( \frac{\lambda \left( (1-\hat{p}) (N+1+b) -N+j \right) }{N+1+b} \right. \nonumber \\
   && \quad \left. +\frac{(1-\lambda) (1-\hat{p}) (b-1)}{N+b} \right).
\label{Substitutionmodified}
\end{eqnarray}
By following the rest of the derivations, it can be seen that conclusion of the theorem still holds.	

\section{Proof of Theorem \ref{MACConvergence}}
\label{ProofofMACConvergence}
First, according to Theorem \ref{monotonicity}, if $b$ is chosen to satisfy $b>\max\{1, x^*-\gamma_{\epsilon_v}\}$, $\frac{dq^*_v(\hat{p})}{d\hat{p}}>0$ holds with strict inequality for $\hat{p}\in (0, p_{\max})$. According to Theorem \ref{PartialMonotonicity}, $q_v(p\mbox{\boldmath$1$}, K)$ is non-increasing in $\hat{p}$ for any given number of users $K$. Therefore, if $K\ge J_{\epsilon_v}$, $q_v^*(\hat{p})=q_v(\hat{p}\mbox{\boldmath$1$}, K)$ should have a unique solution at $\hat{p}=\frac{x^*}{K+b}$. If $K< J_{\epsilon_v}$ on the other hand, we must have $q_v(\hat{p}\mbox{\boldmath$1$}, K)> q_v^*(\hat{p})$ for all $\hat{p}\in [0, p_{\max})$. Consequently, the proposed MAC algorithm should possess a unique equilibrium at $\mbox{\boldmath $p$}^*=p^*\mbox{\boldmath$1$}=\min \{p_{\max},\frac{x^*}{K+b}\}\mbox{\boldmath$1$}$.

Second, consider an arbitrary $\hat{p}<p_{\max}$, which implies that the corresponding $\hat{K}$ should satisfy $\hat{K}>J_{\epsilon_v}$. According to (\ref{Substitution}) and (\ref{derivative}), we have
\begin{eqnarray}
&& \frac{d q_v^*(\hat{p})}{d \hat{p}} \ge \frac{\hat{p}}{p_N - p_{N+1}} \left( {N \atop J_{\epsilon_v}} \right)  \nonumber \\
	&&  \quad \times \hat{p}^{J_{\epsilon_v}} (1-\hat{p})^{N-J_{\epsilon_v}-1}(C_{vJ_{\epsilon_v}} - C_{v(J_{\epsilon_v}+1)}) \nonumber \\
	&& \times \left( \frac{\lambda (b-x^*+J_{\epsilon_v})}{N+1+b} + \frac{(1-\lambda)(1-\hat{p})(b-1)}{N+b} \right). \nonumber \\
	\label{boundderivative}
\end{eqnarray}
It can be seen that the right hand side of (\ref{boundderivative}) has a positive limit as $\hat{p}$ is taken to zero. Therefore, we can find two positive constants $\epsilon_0, \epsilon_1>0$, such that $\frac{d q_v^*(\hat{p})}{d \hat{p}} \ge \epsilon_0 >0$ for all $\hat{p}\le \epsilon_1$. In the meantime, when $\epsilon_1 \le \hat{p}< p_{\max}$, because $b>\max\{1, x^*-\gamma_{\epsilon_v}\}$, we can find another positive constant $\epsilon_2>0$, such that the right hand side of (\ref{boundderivative}) is larger than or equal to $\epsilon_2$. Consequently, there exists a positive constant $\epsilon=\min\{\epsilon_0, \epsilon_2\}$, such that $\frac{d q_v^*(\hat{p})}{d \hat{p}} \ge \epsilon >0$ for all $\hat{p}<p_{\max}$.

Third, let $q^{*-1}_v(.)$ be the inverse function of $q^*_v(p)$. For any given transmission probability vector $\mbox{\boldmath $p$}$, transmission probability target $\hat{p}$ is obtained by
\begin{equation}
\hat{p}=q^{*-1}_v(q_v)=q^{*-1}_v(q_v(\mbox{\boldmath $p$},K)).
\label{QvInverse}
\end{equation}
Because $\frac{d q_v^*(\hat{p})}{d \hat{p}} \ge \epsilon >0$ for all $\hat{p}<p_{\max}$, there must exist a constant $K_{l_1}>0$ such that
\begin{equation}
\|\hat{p}_1-\hat{p}_2\|\le K_{l_1}\|q_{v1}-q_{v2}\|,
\label{Lipschitz1}
\end{equation}
for all $\hat{p}_1=q^{*-1}_v(q_{v1})$ and $\hat{p}_2=q^{*-1}_v(q_{v2})$. Furthermore, because $q_{v}=q_v(\mbox{\boldmath $p$},K)$ is Lipschitz continuous in $\mbox{\boldmath $p$}$, for all $q_{v1}=q_v(\mbox{\boldmath $p$}_1,K)$ and $q_{v2}=q_v(\mbox{\boldmath $p$}_2,K)$, there exists a constant $K_{l_2}>0$ to satisfy
\begin{equation}
\|q_{v1}-q_{v2}\|\le K_{l_2}\|\mbox{\boldmath $p$}_1-\mbox{\boldmath $p$}_2\|.
\label{Lipschitz2}
\end{equation}
From (\ref{Lipschitz1}) and (\ref{Lipschitz2}), for all $\hat{p}_1=q^{*-1}_v(q_v(\mbox{\boldmath $p$}_1,K))$ and $\hat{p}_2=q^{*-1}_v(q_v(\mbox{\boldmath $p$}_2,K))$, we have
\begin{equation}
\|\hat{p}_1-\hat{p}_2\|\le K_{l_1} K_{l_2} \|\mbox{\boldmath $p$}_1-\mbox{\boldmath $p$}_2\|.
\label{Lipschitz3}
\end{equation}
This implies that the probability target function given in (\ref{QvInverse}) satisfies the Lipschitz condition given in Condition \ref{lipschitz}.

Finally, when the system is noisy, the receiver can choose to measure $q_v$ over an extended number of time slots, namely increasing the value of $Q$ introduced in Step \ref{repeat} of the distributed MAC algorithm. If users maintain their transmission probabilities during the $Q$ time slots, by assumption, an increased value of $Q$ can reduce the potential measurement (or estimation) bias in the system arbitrarily close to zero. Therefore the bias condition given in Condition \ref{BiasCondition} is also satisfied. Consequently, convergence of the proposed distributed MAC algorithm is supported by Theorems \ref{strongconvergence} and \ref{weakconvergence}.

\section{Proof of Theorem \ref{onestep}}
\label{Proofofonestep}
According to the two-step approach, $q_v$ is interpreted by $q_v=(1-p_k)q_k+p_kd^*(\breve{p})$. When $\breve{p}\ge p_k$, we should either have $q_k=q^*(\breve{p})$ when $\breve{p}<p_{\max}$, or $q_k\ge q^*(\breve{p})$ when $\breve{p}=p_{\max}$. Therefore
\begin{eqnarray}
&& q_v=(1-p_k)q_k+p_k d^*(\breve{p}) \nonumber \\
&& \ge (1-p_k)q^*(\breve{p})+p_k d^*(\breve{p}) \nonumber \\
&& =q^*(\breve{p})-p_k(q^*(\breve{p})-d^*(\breve{p})) \nonumber \\
&& \ge q^*(\breve{p})-\breve{p}(q^*(\breve{p})-d^*(\breve{p})) \nonumber \\
&& =q^*_v(\breve{p}),
\end{eqnarray}
where the last inequality is due to the fact that $q^*(\breve{p})-d^*(\breve{p})\ge 0$ should always hold.

Because $b\ge \max\{1, x^*-\gamma_{\epsilon_v}\}$, according to Theorem \ref{monotonicity}, $q^*_v(\hat{p})$ is non-decreasing in $\hat{p}$. Therefore, if $q_v>q^*_v(p_{\max})$, we have $\hat{p}=p_{\max} \ge p_k$. Otherwise, we have
\begin{equation}
q^*_v(\hat{p})=q_v\ge q^*_v(\breve{p}) \ge q^*_v(p_k).
\end{equation}	
This also implies that we $\hat{p}\ge p_k$.

Similarly, when $\breve{p}\le p_k$, it can be shown that the two-step approach will yield $\hat{p}\le p_k$.

\section*{Acknowledgement}
This work was supported by the National Science Foundation under Grants CCF-1420608 and CNS-1618960. Any opinions, findings, and conclusions or recommendations expressed in this paper are those of the authors and do not necessarily reflect the views of the National Science Foundation.




\end{document}